\documentclass[]{spie} 

\usepackage{amsmath,amsfonts,amssymb}
\usepackage{graphicx}
\usepackage[colorlinks=true, allcolors=blue]{hyperref}
\usepackage{bm}
\usepackage{float}
\newcommand{\vect}[1]{\boldsymbol{\mathbf{#1}}}

\title{Real-time estimation and correction of quasi-static aberrations in ground-based high contrast imaging systems with high frame-rates}

\author[a,b]{Alexander T. Rodack}
\author[b]{Jared R. Males}
\author[a,b,c,d]{Olivier Guyon}
\author[e]{Benjamin A. Mazin}
\author[f]{Michael P. Fitzgerald}
\author[g,h]{Dimitri Mawet}
\affil[a]{University of Arizona, College of Optical Sciences, Tucson, Arizona, United States}
\affil[b]{University of Arizona, Steward Observatory, Tucson, Arizona, United States}
\affil[c]{National Institutes of Natural Sciences, Subaru Telescope, National Observatory of Japan, Hilo, Hawaii, United States}
\affil[d]{National Institutes of Natural Sciences, Astrobiology Center, Mitaka, Japan}
\affil[e]{University of California, Santa Barbara, Dept. of Physics, Santa Barbara, California, United States}
\affil[f]{University of California, Los Angeles, Dept. of Physics and Astronomy, Pasadena, California, United States}
\affil[g]{California Institute of Technology, Dept. of Astronomy, Pasadena, California, United States}
\affil[h]{NASA Jet Propulsion Lab, Pasadena, California, United States}

\authorinfo{}

\begin{document} 
\maketitle

\begin{abstract}
The success of ground-based, high contrast imaging for the detection of exoplanets in part depends on the ability to differentiate between quasi-static speckles caused by aberrations not corrected by adaptive optics (AO) systems, known as “non-common path aberrations” (NCPAs), and the planet intensity signal. Frazin (ApJ, 2013) introduced a post-processing algorithm demonstrating that simultaneous millisecond exposures in the science camera and wavefront sensor (WFS) can be used with a statistical inference procedure to determine both the series expanded NCPA coefficients and the planetary signal. We demonstrate, via simulation, that using this algorithm in a closed-loop AO system, real-time estimation and correction of the quasi-static NCPA is possible without separate deformable mirror (DM) probes. Thus the use of this technique allows for the removal of the quasi-static speckles that can be mistaken for planetary signals without the need for new optical hardware, improving the efficiency of ground-based exoplanet detection. In our simulations, we explore the behavior of the Frazin Algorithm (FA) and the dependence of its convergence to an accurate estimate on factors such as Strehl ratio, NCPA strength, and number of algorithm search basis functions. We then apply this knowledge to simulate running the algorithm in real-time in a nearly ideal setting. We then discuss adaptations that can be made to the algorithm to improve its real-time performance, and show their efficacy in simulation. A final simulation tests the technique's resilience against imperfect knowledge of the AO residual phase, motivating an analysis of the feasibility of using this technique in a real closed-loop Extreme AO system such as SCExAO or MagAO-X, in terms of computational complexity and the accuracy of the estimated quasi-static NCPA correction.
\end{abstract}


\section{INTRODUCTION}
\label{sec:intro} 
Eliminating quasi-static non-common path aberrations (NCPA) not seen by the AO system is an important step to improving the capability of ground-based, high contrast imaging systems to detect exoplanets. In Frazin 2013, \cite{Frazin767} the author proposed a means to post-process millisecond science camera exposures simultaneously obtained with WFS measurements of the AO residual phase ({$\phi_r$}) to estimate any quasi-static NCPA phase ({$\phi_u$}) upstream of the coronagraph in a high contrast imaging system, as well as any present planetary emission in the science frames. To review, the Frazin Algorithm (FA) is advantageous because of its exploitation of the fact that the AO residual phase provides a new phase in the pupil plane at each exposure, and provides a statistically independent phase screen which modulates the quasi-static speckles in a new way every ``atmospheric clearing time,'' {$\tau_c$} (Macintosh et al. 2005) \cite{Macintosh2005}. This means that the rapidly changing AO residual speckles can be used as a probe to estimate the quasi-static NCPA because with each passing millisecond, more diversity in the observations is achieved. In equation form, the science camera intensity can be written as:
\begin{equation}
\label{eqn: FAIntensity}
I(\vect{\rho},t) = u_{\bullet}^2 i_p(\vect{\rho},t) + \mathcal{A}(\vect{\rho},t) + \vect{a}^{\dagger}\vect{b}(\vect{\rho},t) + \vect{b}^{\dagger}(\vect{\rho},t)\vect{a} + \vect{a}^{\dagger}\mathbf{C}(\vect{\rho},t)\vect{a} \, ,
\end{equation}
where $\vect{\rho}$ is a vector of pixel locations, $u_{\bullet}$ is the planet field amplitude, $u_{\bullet}^2 i_p$ is the planetary point spread function (PSF), $\vect{a}$ is a vector of quasi-static aberration coefficients, $\mathcal{A}$ is the intensity only depending on the AO residual speckles ($\phi_r$), $\mathbf{C}$ depends on the quasi-static aberration ($\phi_u$) as modulated by $\phi_r$, and $\vect{b}$ depends on the mixing of both $\phi_r$ and $\phi_u$. $\phi_u$ is then decomposed into a ``search'' basis set with no orthogonality restriction as:
\begin{equation}
\label{eqn: FAphiu}
\phi_u(\vect{r}) = \sum_{k=1}^Ka_k\psi_k(\vect{r}) \, ,
\end{equation}
where $a_k$ are the individual elements of $\vect{a}$ from above, and $\psi_k$ are the functions in the search basis the algorithm fits the quasi-static NCPA to. Considering N pixel locations $\vect{\rho} = \left\lbrace \rho_1, ..., \rho_N\right\rbrace$ in a single exposure from the science camera, $I(\vect{\rho},t_i)$, one desires to know if there is planet emission, and T total millisecond exposures synchronized with WFS measurements of $\phi_r$,  $\mathcal{A}(\vect{\rho},t_i)$, $\vect{b}(\vect{\rho},t_i)$, and $\mathbf{C}(\vect{\rho},t_i)$ are computed using FFTs following the equations given by Frazin, \cite{Frazin767} and the following linear system model:

\begin{equation}
\label{eqn: LSmodel}
I(\vect{\rho},t) - \mathcal{A}(\vect{\rho},t) = u_{\bullet}^2 i_p(\vect{\rho},t) + \vect{a}^{\dagger}\vect{b}(\vect{\rho},t) + \vect{b}^{\dagger}(\vect{\rho},t)\vect{a} + \vect{a}^{\dagger}\mathbf{C}(\vect{\rho},t)\vect{a} \, ,
\end{equation}
is used to set up an inverse problem as:
\begin{equation}
\label{eqn: FALS}
\vect{y} = \mathbf{H}\vect{x} \, ,
\end{equation}
where
\begin{equation}
\label{eqn: FAy}
\vect{y} = \left[\vect{y_1};\, ...\,;\, \vect{y_T}\right] \, , \{\vect{y_i}\} = I(\vect{\rho_n},t_i) - \mathcal{A}(\vect{\rho_n},t_i) \, ,
\end{equation}
\begin{equation}
\label{eqn: FAH}
\mathbf{H} = \left[\mathbf{H_1};\, ... \,;\,\mathbf{H_T}\right] \, , \{\mathbf{H_i}\} = \left[i_p(\vect{\rho_n},t_i)\qquad \vect{b}^T(\vect{\rho_n},t_i)\qquad \vect{b}^{\dagger}(\vect{\rho_n},t_i)\qquad \vect{c}^{\dagger}(\vect{\rho_n},t_i)\right]\, ,
\end{equation}
and
\begin{equation}
\label{eqn: FAx}
\vect{x} = \left[u_{\bullet}^2; \vect{a}; \vect{a}^*; \{\vect{a}_k\vect{a_l}^*\}\right]
\end{equation}
$\vect{x}$ is then solved for to produce an estimate of the planetary field amplitude and the coefficients of the search basis set functions. The methods described below aim to take these equations and compute them in real-time using data streams from a science camera and WFS, as opposed to post-processing. Achieving this would allow for real-time estimations of the slowly evolving NCPA to be generated, and applied to a DM or DMs to correct for them, thus eliminating the quasi-static speckles one might mistake for a planetary signal.

\section{VERIFYING THE FRAZIN ALGORITHM}
\label{sec:PFA}
\begin{figure}
\centering
\includegraphics[scale=0.25]{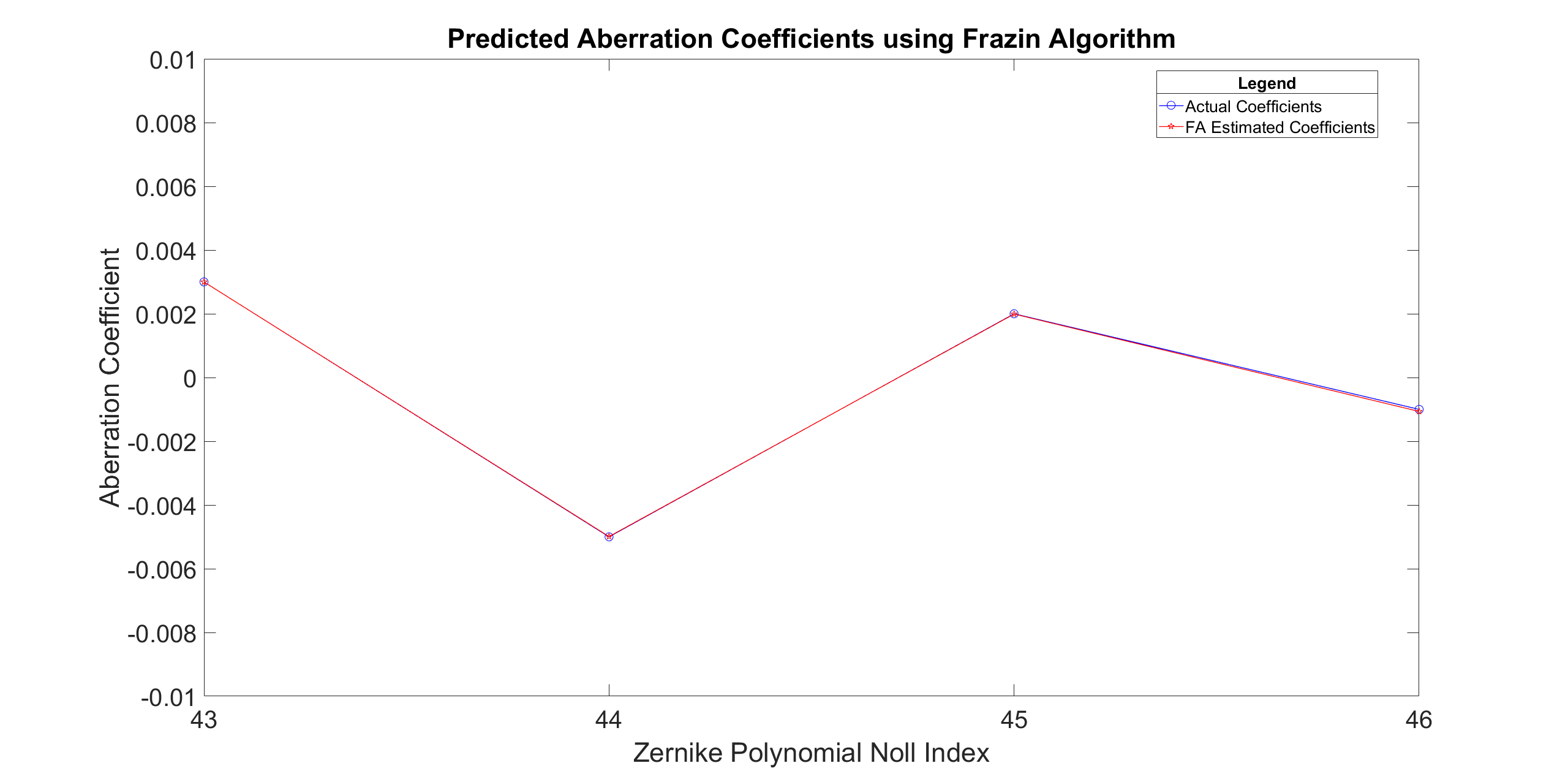}
\caption{Predicted and actual NCPA basis coefficients using the Frazin Algorithm with 500 one millisecond exposures.}
\label{fig: CoeffAccuracy}
\end{figure}

\begin{figure}[t]
\centering
\includegraphics[scale=0.33]{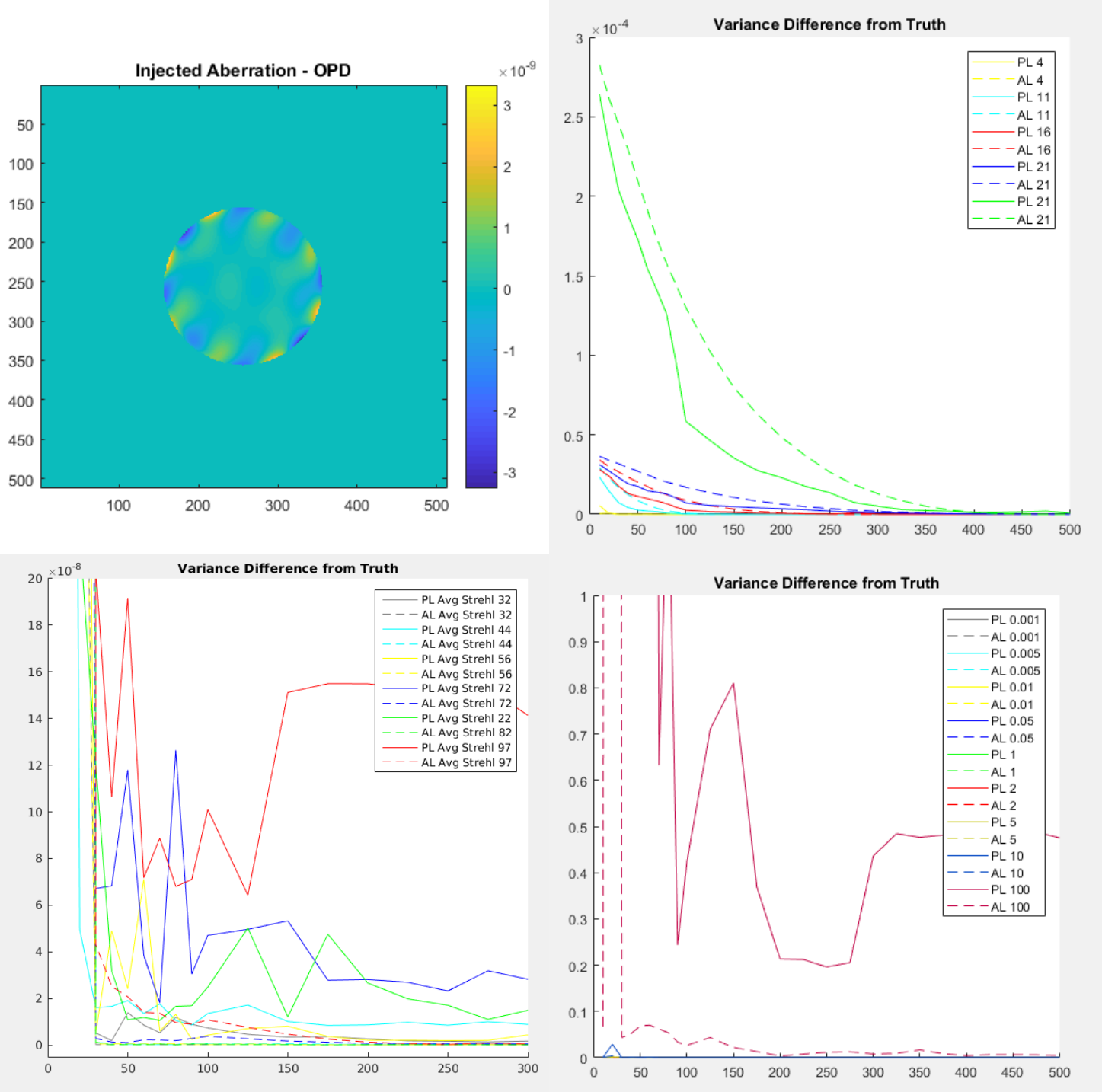}
\caption{At each exposure measured, the Frazin Algorithm is used to generate an estimate of the NCPA using a single pixel location (PL) for the estimate, and using all the pixels in the search region (AL). This estimate is then subtracted from the true NCPA, and the variance across the pupil is measured and plotted versus the exposure number. When the variance difference between the truth NCPA and the estimated NCPA is minimized, the algorithm has converged. The top left frame shows the form of the injected phase NCPA in OPD. The top right frame shows the variance difference vs. number of exposures in the calculation vs. the number of basis functions in the search basis (numbered in the legend). In all cases except the green lines, an aberration basis set of 4 functions is used. The green lines use 15 functions in the aberration basis set. The bottom left frame shows the variance difference vs. number of exposures in the calculation vs. average Strehl ratio. The bottom right frame shows the variance difference vs. number of exposures in the calculation vs. strength of the NCPA (scale factors applied shown in legend, scale = 1 is what is shown in the top left frame).
}
\label{fig: convergencerates}
\end{figure}

In order to better understand the feasibility of running the Frazin Algorithm (FA) in real-time, a simulation is set up in order to both verify that the algorithm works as described, and explore its parameter space. Using a simple Fraunhofer diffraction model, we perform plane-to-plane propagations between elements of a kHz AO system which includes an ideal WFS, an ideal coronagraph, and a noiseless detector. A model of the atmosphere is constructed using AtmosphericTurbulenceSimul (see \url{https://github.com/oguyon/AtmosphericTurbulenceSimul}). This models a seven layer atmosphere, with each layer at various altitudes and wind vectors, and uses Fresnel Propagation between the layers to collapse the effects into a single phase screen with an average seeing of 0.65 arcseconds at 500nm, at 1 millisecond time steps. Only the phase effects are used in this simulation, leaving amplitude effects for a later study, knowing the algorithm is designed to handle them. A bright planewave (1e10 photons per millisecond per square meter) to model starlight is sent through the atmosphere model, and is incident upon an eight meter diameter telescope pupil. Here, an ideal WFS measurement is taken to measure $\phi_r$, and then a phase only NCPA ($\phi_u$) that consists of the sum the high order Zernike polynomials Z43-Z46 (known as the aberration basis; pictured in units of meters of optical path difference (OPD) in Figure \ref{fig: convergencerates} top left) is injected. The ideal coronagraph is then applied, and Fourier Optics is used to compute the star PSF at the ideal science camera. A second, off-axis planewave (100 photons per millisecond per square meter for a contrast ratio of 1e-6 to the star) is then used to compute a planetary PSF, which is summed with the star PSF to obtain the final science camera frame. At each exposure, the components of $\vect{y_i}$ and $\mathbf{H_i}$ are computed for each pixel in a 13x13 pixel square centered at the middle of the science camera data known as the search region. These components are stored to be used to solve for $\vect{x}$ once 500 exposures are simulated. This simulation is designed as the ``control'' case for exploring the dependence of the FA on some important parameters. The AO system is tweaked to provide an average Strehl across the time series of 0.55, with the algorithm search basis being the same Zernike polynomials in the aberration basis. The results for the predicted basis coefficients can be seen in Figure \ref{fig: CoeffAccuracy}. The estimation of the planet field amplitude is 99.964 photons per ms per square meter, or an accuracy of 99.9\%.

With the algorithm performing as described for the Control case above, simulations are run to explore the dependency of the convergence rate and accuracy of the algorithm's estimates on several parameters, including the number of basis functions included in the search basis, the Strehl ratio, and the strength of the NCPA. The behavior of the convergence rate is important because it informs us of the requirements for the AO system as well as the control system employing the FA to correct NCPAs. Using the same simulation setup as described above unless otherwise noted, individual factors are run at a range of values, and the variance of the difference between the true injected NCPA and the estimated NCPA is measured every 25 exposures. The point of convergence is defined where this variance difference is minimized and stable, and the accuracy of the estimates is judged as the value of the variance difference at this point. Two methods  of generating the estimates are also examined: (1) using the estimate found from a single pixel location in the search region (denoted as PL in the plots) and (2) using the average of the estimates found from each of the 169 pixel locations in the search region (denoted as AL).

The first dependence that is analyzed is the number of functions included in the search basis. Because the number of unknowns increases with the number of search basis functions, more diversity is required to obtain an accurate estimate of the NCPA. The expected behavior is thus that convergence will require more exposures as the number of functions increases. The convergence plots can be seen in the top right of Figure \ref{fig: convergencerates}. As expected, the convergence rate slows as the number of search basis functions increases.  A second test where the injected NCPA is the linear sum of 15 Zernike polynomials, and 21 Zernikes are used in the search basis is performed, and it is seen in the green line in the same figure frame. The purpose of this test is to examine what happens if the NCPA is more complicated. It is noted that the convergence rate increases by 100 exposures as compared to the less complicated NCPA, but that the variance difference starts much higher and still converges to a similar value. By comparing the two methods of generating the estimated coefficients, it is seen in this case that the AL convergence curve is smoother and more stable than the PL curve, but that the convergence rate is slightly slower.

The second dependence that is analyzed is the average Strehl ratio over the time series. The Strehl ratio is manipulated in the simulation by restricting the spatial frequencies the AO system is capable of correcting rather than weakening the strength of the phase screen. The expected behavior is that as the Strehl ratio increases (better AO system performance), the convergence rate will decrease slightly. With a lower Strehl ratio, the AO residual speckles produced by $\phi_r$ are stronger, and act as better probes modulating $\phi_u$. A higher Strehl ratio reduces $\phi_r$, even possibly to the point of the NCPA becoming the dominant aberration present in the wavefront, reducing the efficiency of the FA. The convergence plots for this analysis can be seen in the lower right of Figure \ref{fig: convergencerates}. Again, the expected behavior is present. However, a vast difference in the convergence between the PL and AL methods is also present. The AL method curves remain smoother, but converge very quickly to lower variance difference values than the PL method curves. The PL method in this case seems to become very unreliable, suggesting that an individual pixel estimate requires more time to converge to the true value with higher Strehl, but the average of all the individual pixel estimates in the search region maintains the precision we expect to see in fewer exposures.

The final dependence that is examined is the strength of the NCPA that is being looked for. In this case, a scale factor is applied to the injected NCPA in the simulations. The scale factor of one is what is pictured in the top left of Figure \ref{fig: convergencerates}. The other scale factors used can be found in the legend of the bottom right frame of Figure \ref{fig: convergencerates}. The expected behavior is that the FA will perform better for smaller aberrations, and degrade quickly as the scale becomes larger because the construction of the algorithm uses only up through the quadratic terms of the Taylor expansion. What is seen in Figure \ref{fig: convergencerates} shows this to be the case. All of the small aberrations (scale less than 100) converge on a scale that is orders of magnitude faster than the largest aberration analyzed. Using a scale factor of 100 represents an aberration strength on par with one wavelength in OPD, which falls outside of the small aberration assumption used to justify throwing away the terms of the expansion larger than the quadratic term. The same behavior that is observed comparing the PL and AL methods in the case of the Strehl ratio examination is present in this test as well.

Having now examined numerous different simulated experiments in probing the algorithm, and finding that it converges and provides very accurate estimates in all the cases it is expected to, we confirm the validity of the FA. It is also clear that the AL method should be used for future work with the FA in order to obtain estimates of the aberration coefficients and planet amplitude with the quickest, most stable convergence. With a greater understanding of the FA and its behavior, we move on to simulate its use in a real-time, integrator control loop to correct the NCPA.

\section{SIMULATING THE FRAZIN ALGORITHM IN REAL-TIME}
\label{sec:FA-RT}
In order to simulate the FA in real-time, several assumptions are necessary. Again, the ideal coronagraph, ideal WFS, ideal science camera, and the same phase only model of the atmosphere are chosen, but in order to truly demonstrate the real-time capability, a new static NCPA is constructed. This NCPA is chosen to be a phase only cosine mode to create a vertical speckle pair in the science camera image plane, that is perturbed by the same Zernikes (Z43-Z46) as used before. The cosine mode is chosen to place the speckle pair such that one of the speckles lies directly on top of an injected planet PSF. The AO system is manipulated to maintain a very high averaged Strehl ratio of about 0.95 across the length of time the simulation is run. This is done so that the ideal coronagraph model performs better, allowing the planet with a contrast ratio of $10^{-3}$ to become visible in the science camera image plane if the NCPA is corrected. The simulation computes the intensity at the science camera image plane for each millisecond exposure and the calculations required by the FA before moving on to the next exposure. This means an assumption is made that the computer running the FA is capable of finding each exposure's contribution to $\vect{y}$ and $\vect{H}$ as fast as exposures are taken. This also makes the assumption that the WFS readout of $\phi_r$ is exactly simultaneous with the science camera readout. While both of these assumptions are easily violated in real life observations, they are fine to make in this case as a proof of concept that the FA can be run in real-time. They both, however, inform the feasibility of the technique overall, and will be examined in a later study.

\subsection{RUNNING THE BASE ALGORITHM IN REAL-TIME}
\label{sec: BART}
To implement the FA as a real-time control algorithm, the standard AO control technique of an integrator is chosen due to its proven robustness in the field. Equation \ref{eqn: FA integrator} displays the formulation of the FA real-time loop integrator: 

\begin{equation}
\label{eqn: FA integrator}
(\hat{\psi}_{DM})_n = (\hat{\psi}_{DM})_{n-1} + g_{FA} \frac{(\hat{\phi}_u)_n}{k} \, ,
\end{equation}
where $\hat{\psi}_{DM}$ is the DM OPD command sent at estimate iteration number $n$, $g_{FA}$ is the FA loop gain, $k$ is the wavenumber, and $\hat{\phi}_u$ is the current estimate of the NCPA phase. This leads to a batch estimation scheme where millisecond exposures are gathered for the time, $T_{c}$ milliseconds, it takes for the FA to reach convergence. At each exposure in $T_c$, $\vect{y}$ and $\vect{H}$ are updated, and once $T_c$ exposures are measured, Equation \ref{eqn: FALS} is solved to generate $\hat{\phi}_u$ from the estimated $\vect{a}$ coefficients. Equation \ref{eqn: FA integrator} is then used to update the DM, $\vect{y}$ and $\vect{H}$ are cleared, and a new FA iteration begins.The next iteration thus is responsible for detecting the residual NCPA left, leading to a closed-loop system to remove the static NCPA.

\begin{figure}
\centering
\includegraphics[scale=0.25]{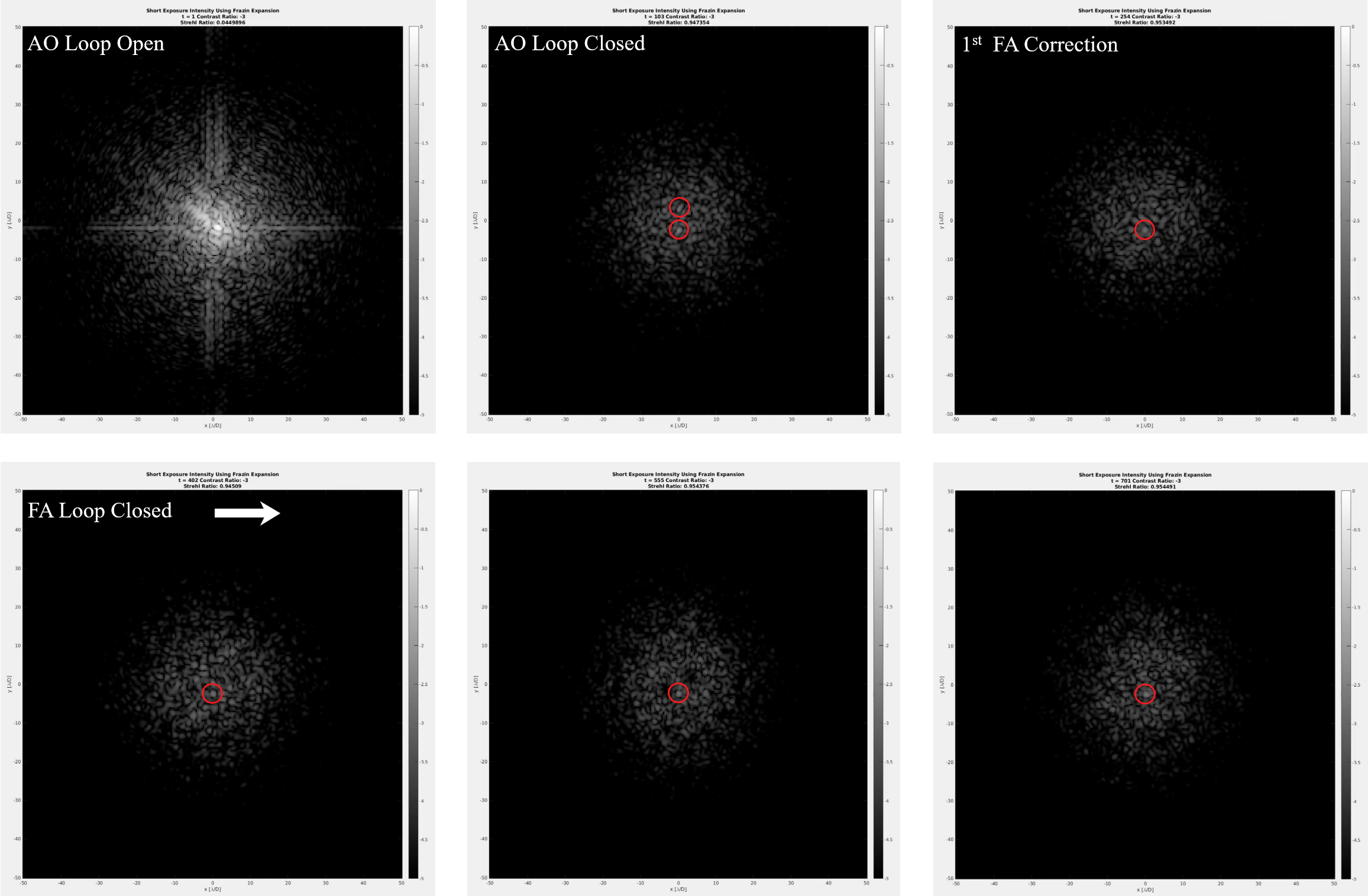}
\caption{Frames from various points in the real-time simulation of the FA showing the removal of the speckles created by a static NCPA.}
\label{fig: IntensityFrames}
\end{figure}

Using the knowledge from the parameter space exploration of Section \ref{sec:PFA}, the real-time simulation is set up to help minimize the number of exposures required for the algorithm to converge. Because it is in simulation, the option to match the algorithm search basis to the aberration basis is chosen to allow for the fastest convergence of the first estimate. $T_c$ is set at 150 however to allow for enough time for all estimates after the first iteration to have more time to fit the residual NCPA to the five search basis functions as best as possible. The simulation runs for a total of 1050 millisecond exposures, where the AO system is closed at the 100th exposure, where the FA control loop starts running. Every 150 exposures, Equation \ref{eqn: FA integrator} is called, and a DM update is sent. $g_{FA}$ is chosen to be 0.8 in order to slow the correction and ensure its stability. This was determined through experimenting with different gain values and choosing one that had good performance. Figure \ref{fig: IntensityFrames} shows frames out of the intensity time series corresponding to the AO loop being open and closed, and then when successive iterations of the FA control loop having been applied. Once the AO loop closes, the NCPA speckle pair is clearly visible (shown in red circles). After the first correction generated by the FA, the top speckle of the pair is largely removed, leaving behind the injected planet PSF (in the single red circle). After the second iteration, the top speckle is completely removed, leaving only the planet PSF visible. At this point, the FA loop is closed, and continues to keep the static NCPA controlled.

\begin{figure}
\centering
\includegraphics[scale=0.50]{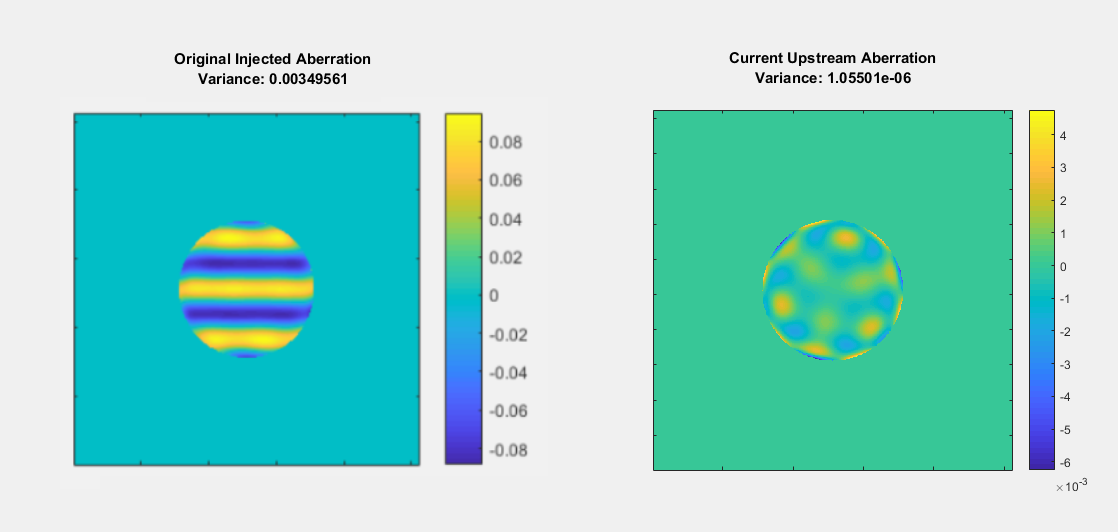}
\caption{The injected static NCPA and a residual NCPA after the FA control loop is closed are shown in phase. The variance across the pupil in which the aberration lies is shown above the figure.}
\label{fig: VarianceReduction}
\end{figure}

Figure \ref{fig: VarianceReduction} shows the initial static NCPA in phase on the left, and the residual NCPA at the end of the simulation in phase on the right. Looking at the form of the residual, the cosine mode is no longer present, but there are low order effects that come from the algorithm search basis looking for the cosine mode and high-order Zernike polynomials to correct only. This residual is expected because the search basis can only do so well at matching the residual, and thus only provides small corrections once the FA loop is closed. The variance across the pupil is measured and displayed above each image, and is used as the merit value for the effectiveness of the correction. The variance has decreased by just over three orders of magnitude, and the peak to valley decreases by about 95\%. This confirms the FA's ability to be used as real-time control algorithm running underneath a ground-based high contrast imaging kHz AO system in ideal settings. 

\subsection{ADAPTING THE ALGORITHM: THE REAL-TIME FRAZIN ALGORITHM}
\label{sec:AA}
Although the FA is now shown to work in a real-time control algorithm, some consideration must be given to its computational efficiency. Because the simulation in Section \ref{sec: BART} is computed under the assumptions given, it stands to reason that some adaptations to the algorithm can be performed so that real life use will not so easily violate them. Using the FA in real-time in a control algorithm means that estimating the planet field amplitude can be thought of initially as extra information that does not inform the prediction of the aberration coefficients. With that in mind, the linear system model seen in Equation \ref{eqn: LSmodel} is modified to no longer produce these estimates, creating the Real-Time Frazin Algorithm (RTFA):

\begin{equation}
\label{eqn: AFA}
I(\vect{\rho},t) - \mathcal{A}(\vect{\rho},t) - u_{\bullet}^2 i_p(\vect{\rho},t) = \vect{a}^{\dagger}\vect{b}(\vect{\rho},t) + \vect{b}^{\dagger}(\vect{\rho},t)\vect{a} + \vect{a}^{\dagger}\mathbf{C}(\vect{\rho},t)\vect{a} \, ,
\end{equation}

\begin{figure}
\centering
\includegraphics[scale=0.45]{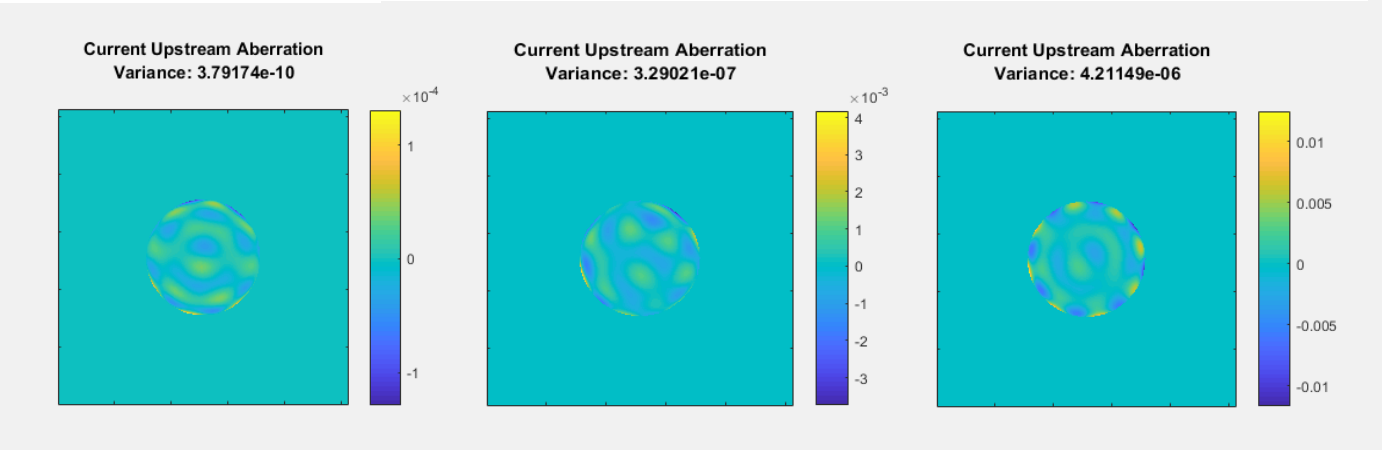}
\caption{The original static NCPA shown in phase, and the residual NCPA remaining after 16 RTFA updates (2500 exposures) in closed loop are shown for various planet brightness levels, with the variance across the pupil shown above the image. Left: No planet present. Middle: Planet at $10^{-3}$ contrast ratio present. Right: Planet at $10^{-2.5}$ contrast ratio present}
\label{fig: RTFA}
\end{figure}

This moves the component of the science camera intensity that comes from a planet from $\vect{H}$ in to $\vect{y}$, shrinking the number of elements in $\vect{H}$ by the number of pixels in the search region for each exposure stored, and the number of elements in $\vect{x}$ by the number of pixels in the search region, without changing the number of elements in $\vect{y}$. With the number of elements dramatically reduced, a well optimized linear solver can be implemented instead of doing a least-squares, minimum norm matrix inversion calculation to solve for $\vect{x}$ much more efficiently than in the previous form. However, in Equation \ref{eqn: AFA}, the planet intensity component must be treated as a known quantity, which is likely not often the case in observations. To solve this, the assumption that the planet intensity is zero (there is no planet light reaching the science camera) is used. Mathematically, this turns $u_{\bullet}^2 i_p(\vect{\rho},t)$ into a source of pixel location dependent additive noise in $\vect{y}$ when using the RTFA. In order to understand the cases in which this assumption holds and the RTFA still maintains the level of accuracy seen using the FA above, the simulation is run again, using the RTFA linear model in place of the FA model, increasing the number of search basis functions to include eight more Zernike polynomials (Z38-Z50), and changing the brightness of the injected planet while providing the RTFA with no knowledge of its presence. Figure \ref{fig: RTFA} displays the residual NCPA still present after after 16 RTFA loop corrections for 3 cases: No planet present, a planet at a contrast ratio of $10^{-3}$, and a planet at a contrast ratio of $10^{-2.5}$. If there is no planet present, the RTFA does not have any error present in $\vect{y}$ due to the planet intensity term. The variance across the residual NCPA is improved by seven orders of magnitude, which performs much better than is seen in Section \ref{sec: BART}. This is expected for two reasons: the RTFA uses a priori knowledge of the planet in this case rather than needing to estimate it, reducing the error in the estimate of the aberration coefficients; the increased number of search basis functions allows for a more robust fit to the residual NCPA following each correction. If the planet is at a contrast ratio to the star of $10^{-2.5}$, the RTFA control loop reduces the variance across the pupil by a comparable amount to what is seen in Section \ref{sec: BART}. The reduction in performance compared to the no planet case is that the RTFA has additional uncertainty in $\vect{y}$ from a planet that is present for which the algorithm does not know is present. This analysis suggests that, at least under ideal conditions and assumptions of the simulation, the RTFA control loop removes static speckles so long as the star is no less than 300 times brighter than a planet that is present. The planets targeted by modern ground-based, high contrast imaging systems are at least $10^6$ times dimmer than their host stars, which makes the assumption under which the RTFA performs valid. A future study will examine the validity and effects of using the FA to estimate the planetary intensity term along side the RTFA control loop to provide a non-zero input, resolving this assumption for all cases of planetary contrast ratio.

\subsection{SIMULATING THE REAL-TIME ALGORITHM: IMPERFECT AO RESIDUAL KNOWLEDGE}
\label{sec:RTFA-IAORK}

\begin{figure}[t]
\centering
\includegraphics[scale=0.55]{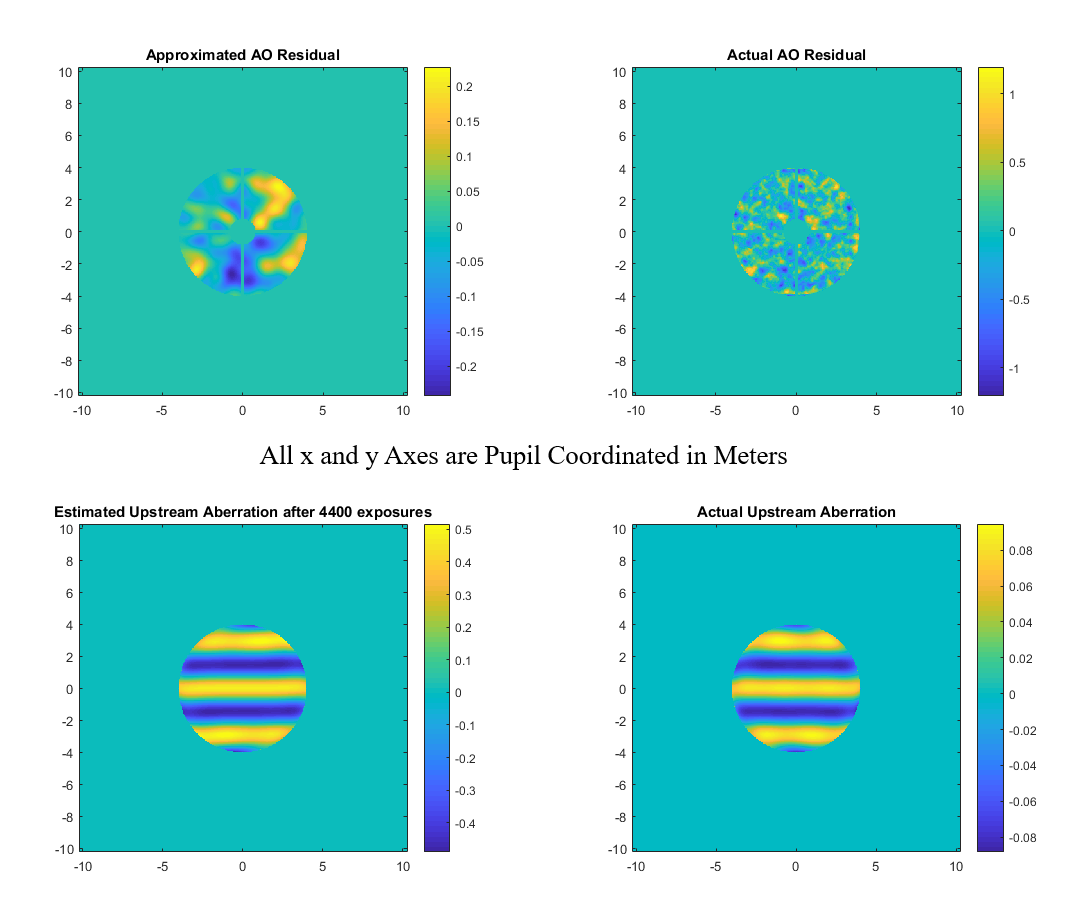}
\caption{Resulting estimation of the NCPA aberration coefficients using the RTFA with a low-order reconstruction of $\phi_r$. Top left: Reconstructed AO residual $\hat{\phi_r}$. Top right: Actual $\phi_r$. Bottom left: Estimated NCPA using RTFA with $\hat{\phi_r}$. Bottom right: True NCPA}
\label{fig: ApproxAORes}
\end{figure}

Using the RTFA in real observations will break the assumptions of using an ideal WFS. Because the algorithm is heavily dependent on knowing $\phi_r$, it is important to examine the effects of using an estimate of it provided by a WFS. Even the best WFS telemetry returns a low-order reconstruction of $\phi_r$ that is missing some high-spatial frequency behavior. The effect of this is more uncertainty that becomes prevalent throughout the calculations of the RTFA. In order to examine the behavior of the RTFA control loop in these circumstances in simulation, a Gaussian kernel is convolved with the ideal WFS measurement already being taken (See Figure \ref{fig: ApproxAORes} top) to produce a representative low-order reconstruction, and fed in to the algorithm as $\phi_r$. It is found that the convergence of the RTFA estimates takes 29 times longer (4400 exposures) and is less stable. In addition, the estimated coefficients relative to one another remains accurate, but an overall scale factor error appears (See Figure \ref{fig: ApproxAORes} bottom). This occurs due to the uncertainty in $\phi_r$ adding uncertainty to the $\vect{H}$ operator that will not be present in the science camera measurement of the total intensity (as the true wavefront propagates through the system to the final focal plane). This discrepancy thus further increases the uncertainty in $\vect{x}$ once it is solved for, effecting the estimated aberration coefficients. The behavior seen is expected, and is very complicated to account for, but a later study is planned to address this and minimize its effects.

\section{CONCLUSIONS AND FUTURE PLANS}
\label{sec:FP}

After this examination, we conclude that the RTFA is a viable technique for estimating and correcting the effects of quasi-static NCPA present in a ground-based, high-contrast kHz AO system. A number of steps to further develop this technique are planned for the coming months. It is clear from Section \ref{sec:RTFA-IAORK} that work needs to be done to further characterize and deal with the propagation of uncertainty and error from using an \vect{H} matrix constructed using an estimate of $\phi_r$. Along side accounting for this, more robustness will be added to the simulation to better represent real-life conditions. Amplitude and phase effects from the atmospheric turbulence and NCPA, noisy detectors, a PIAACMC/vector-vortex coronagraph model, and a quasi-static evolution of the NCPA following predicted real-life speckle lifetimes are among the effects to be included. The computational efficiency of the RTFA will also be examined to get it running as fast as possible. This includes steps such as pre-computing all FFT calculations and storing the results to be called when needed (see Frazin \cite{FrazinPyWFS}), taking advantage of GPU mass parallelization, analyzing if dropping the quadratic terms in the expansions to speed up calculations still allows the aberrations being sought to correct to be estimated correctly, and implementing a scheme to try to reduce the convergence time of the algorithm by estimating any present planetary signal and piping it in to the estimation of the aberration coefficients. After tackling these topics in simulation, the goal is to deploy the technique in a laboratory setting to verify our work and make any final adjustments to the code, and then run it on-sky at Subaru Coronagraphic Extreme Adaptive Optics system (SCExAO) \cite{Lozi2018} and Magellen AO (MagAO). Further in to the future, the RTFA will be modified to run at MagAO-X \cite{Males2018} and SCExAO using MKIDs \cite{Walter2018} as the science detector. Wavelength resolution in the science camera will provide more diversity to the data being supplied to the algorithm, allowing for more robust performance in terms of accuracy and convergence rate.

\appendix

\acknowledgments     
 
This work was supported in part by NSF ATI Award \#1710356 and by NSF MRI Award \#1625441. The authors would also like to thank Dr. Richard A. Frazin for his work developing the base algorithm and his support in our endeavors. 

% References
\bibliography{report} % bibliography data in report.bib
\bibliographystyle{spiebib} % makes bibtex use spiebib.bst

\end{document}